\documentclass[aps,amsmath,amssymb,twocolumn,floatfix,10pt,showpacs]{revtex4-1}

\usepackage{graphicx}
\usepackage{color}

\begin{document}

\title{An eight-fold optical quasicrystal with cold atoms}
\author{Anuradha Jagannathan}
\affiliation{Laboratoire de Physique des Solides, Universit\'e Paris-Sud, 91405 Orsay, France}
\author{Michel Duneau}
\affiliation{CPHT, Ecole Polytechnique, UMR 7644, F-91128 Palaiseau Cedex, France }
\date{\today}
\pacs{  67.85.-d, 61.44.Br, 71.10.Fd}

\begin{abstract}
We propose a means to realize two-dimensional quasiperiodic structures by trapping atoms in an optical potential. The structures have eight-fold symmetry and are closely related to the well-known quasiperiodic octagonal (Ammann-Beenker) tiling. We describe the geometrical properties of the structures  obtained by tuning parameters of the system. We discuss some features of the corresponding tight-binding models, and experiments to probe quantum properties of this optical quasicrystal.
\end{abstract}

\maketitle

\section{Introduction}
In recent years, ultra-cold atoms in optical lattices have provided a way to experimentally realize tight-binding models for bosons or fermions. Tight-binding models, where particles are assumed to be well-described in terms of strongly localized atomic orbitals,  are widely used in condensed matter physics, as they allow to study the electronic states in many types of structures, and help to gain a qualitative understanding of electronic band structure, predict  phase transitions, investigate magnetism, etc. Unlike most solid state materials, cold atom systems can be devised for which the  tight-binding model holds to a good  approximation \cite{bloch}, and for which the parameters of the models can be tuned. We will describe in this paper how one can use the optical potential due to four standing wave laser fields to realize a two dimensional quasiperiodic cold atom quasicrystal having an eight-fold symmetry which is closely related to the well-known octagonal  or Ammann-Beenker tiling \cite{beenker,grun}. We describe the structure and its properties, and the effective tight-binding model to describe fermions or bosons trapped in this optical quasicrystal. This type of system, if realized experimentally, should provide valuable insights into the quantum properties of quasicrystals. An interesting experiment would be to study the quantum dynamics of wave packets prepared at a given energy, as calculations in a number of quasiperiodic models show the possibility of anomalous (super)diffusion with exponents which depend on the initial position of the wave packet.

\section{Cold atoms in a quasiperiodic optical potential \label{sec_atoms}}
We consider atoms trapped in a region where standing waves have been set up using four laser beams oriented at 45$^\circ$ angles in the $xy$ plane and with polarizations perpendicular to this plane. 
For the case of four standing waves, assuming they are all of equal amplitude, with different phase shifts $\phi_n$, the intensity is given by
$I(\vec{r}) =  I_0 \left[\sum_{n=1}^4 \cos(\vec{k}_n.\vec{r}+ \phi_n)\right]^2 $,
\label{Ixy.eq}
where $\vec{r}=(x,y)$ and the four wave vectors are 
\begin{equation}
\vec{k}_n=k(\cos\theta_n,\sin\theta_n) \qquad \qquad \theta_n=\frac{(n-1)\pi}{4}
\label{kvecs.eq}
\end{equation}
with $n=1,..,4$. The dipole potential seen by an atom, $V(\vec{r})$, derives from the AC Stark shift of the atomic levels. In the limit of large detuning $\delta = \omega_L-\omega_{at}$ (where $\omega_L$ is the laser frequency and $\omega_{at}$ is the atomic resonance frequency) the potential is given by
\begin{equation}
V(\vec{r}) =  V_0 \left[\sum_{n=1}^4 \cos(\vec{k}_n.\vec{r}+ \phi_n)\right]^2 
\label{Vxy.eq}
\end{equation}
where $V_0 \approx 3\pi c^2\Gamma I_0/(8\delta \omega^3_{at})$, when the excited state width $\Gamma \ll \vert\delta\vert$ \cite{grimm}.
The potential is thus proportional to the local intensity, and can be positive or negative depending on the sign of $\delta$. The optical potential corresponding to Eq.\ref{Vxy.eq} is an instance of a quasiperiodic function, the theory of which goes back to H. Bohr \cite{bohr} and A. Besicovitch \cite{besic}. Quasiperiodic potentials have been proposed before in the literature. A one dimensional bichromatic lattice was investigated by Deissler et al \cite{deiss}, for interacting bosonic atoms. Quasiperiodic potentials in the plane were realized in \cite{guidoni,sanchez,cetoli}, to study effects of quasiperiodicity in cold atom gases and, incidentally, also in colloidal systems as in \cite{bechinger}.  In this paper we provide the first explicit description of a quasiperiodic structure obtained using an optical potential, which we call an optical quasicrystal (OQ). This structure,  in which atoms are separated by edges of a fixed length, is closely related to the standard octagonal tiling (OT) and its structure factor has an infinite set of Bragg peaks, as expected for a quasicrystal. 

\begin{figure}[!ht]
\centering
\includegraphics[width=140pt]{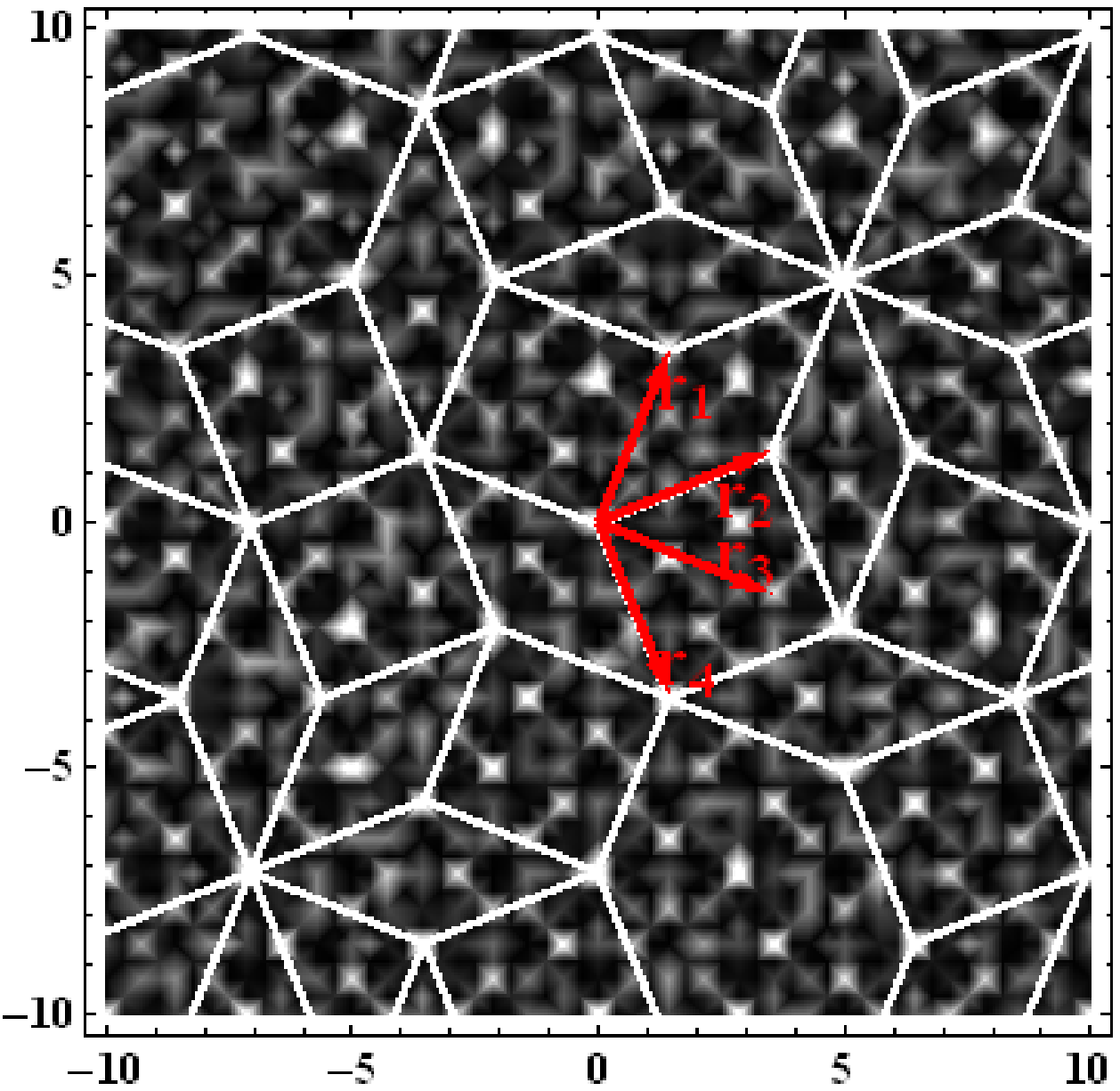}
\includegraphics[width=140pt]{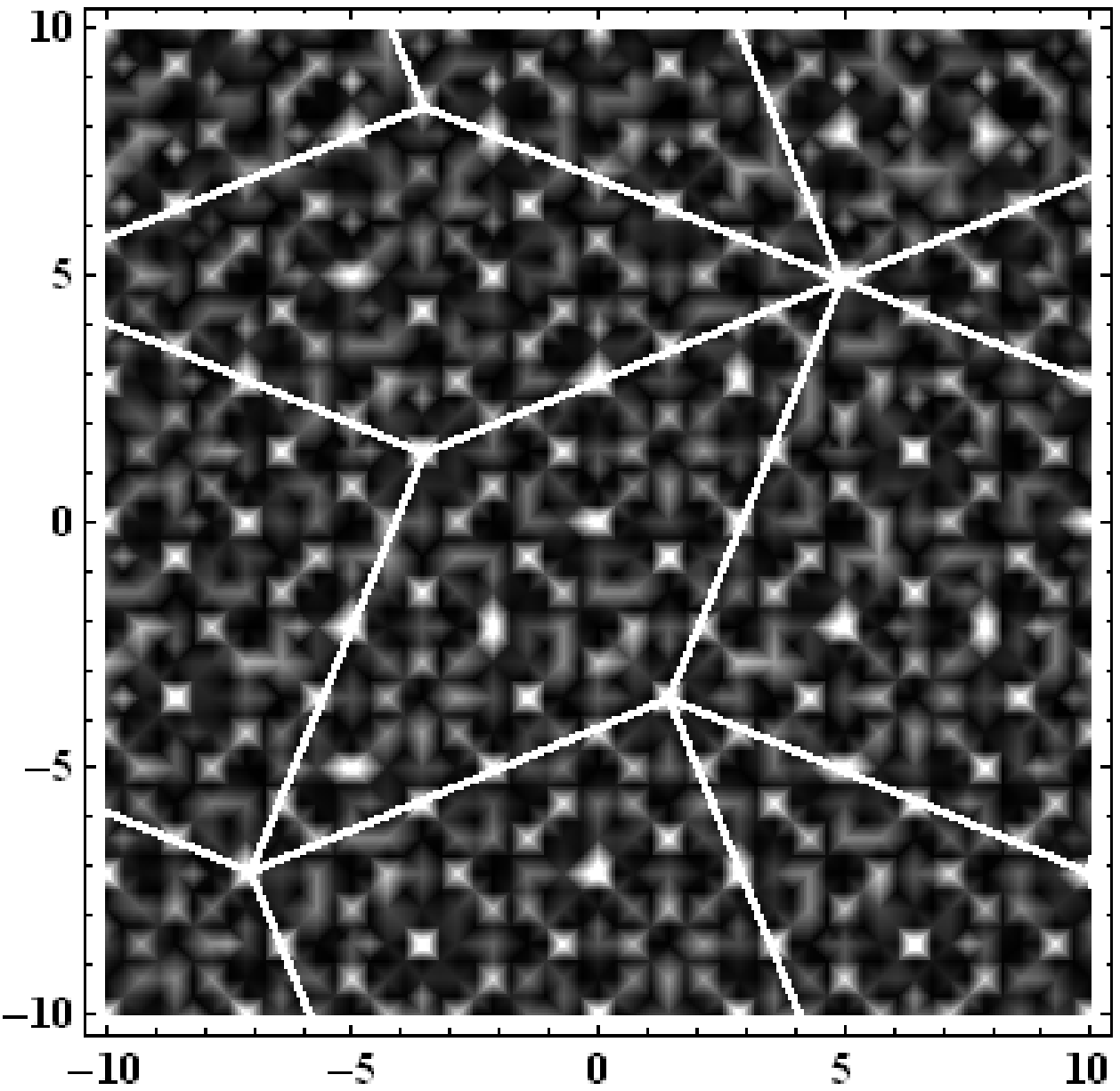}
\caption{Intensity plot of the optical potential in the $xy$ plane, occupied sites, and the resulting tilings for $(I_m-I_c)/I_m= 0.17$ and $0.03$.  The top figure indicates four quasi-lattice vectors $\vec{r}_j$. Distances are indicated in units of $\lambda$. }
\label{one.fig}
\end{figure}

Eq.\ref{Vxy.eq} describes a complex intensity landscape of saddlepoints and local maxima of intensities $0 \leq I \leq I_m\equiv16\vert I_0\vert$. This choice of out-of-plane beam polarization implies a large intensity contrast between the maxima and minima, whereas taking in-plane polarizations would give a smaller contrast ($I_m \approx 6.83 |I_0|$) and therefore less well localized atoms.  We now consider cold atoms confined to the $xy$ plane via a harmonic trap in the $z$ direction, and subjected to this potential. In the limit of a potential with strong variations, the atoms will be localized at certain sites providing the temperature is sufficiently low. In the case of a red-detuned lattice, $\delta < 0$, these sites correspond to the local maxima of $I(\vec{r})$. Depending on the temperature and on the number of available atoms  ( the ``filling" of the optical lattice) the occupied sites are thus those local maxima for which $\vert I(\vec{r}_j)\vert \geq I_c$ where $I_c$ is a cut-off. As the value of cut-off approaches the maximum value $16\vert I_0\vert$, only the largest peaks (corresponding to the lowest energy states) are occupied, and the density of sites decreases. Fig.\ref{one.fig} shows structures obtained for two particular choices of the cut-off.  In each case, the figure shows the occupied sites, and links between them, forming the edges of a tiling. We will show that the four edge-vectors, $\vec{r}_n$, as illustrated in Fig.1a can be expressed in terms of the laser wavelength $\lambda$ and the irrational number, $\alpha =1+\sqrt{2}$, sometimes called the silver mean. 

Figs.\ref{one.fig} show that the patterns appearing in the optical system are very similar to those found in the standard octagonal tiling \cite{octagonal1}, composed of squares and 45$^\circ$ rhombuses. Patterns in the OT repeat quasiperiodically in space, and occur in eight equiprobable orientations. Its Fourier transform (structure factor) has perfect eight-fold symmetry, and comprises a dense set of Dirac delta functions of different intensities, most of which are vanishingly small.  An important property of the octagonal quasiperiodic tiling concerns its invariance under scale changes, called inflations/deflations, of tiles by the factor $\alpha$. These properties are shared by the optical quasicrystal. In the following section, a number of geometrical properties are  derived and explained in terms of a model in four dimensional space.  

\section{Four dimensional model for the optical quasicrystal}
\begin{figure}[!ht]
\centering
\includegraphics[width=240pt]{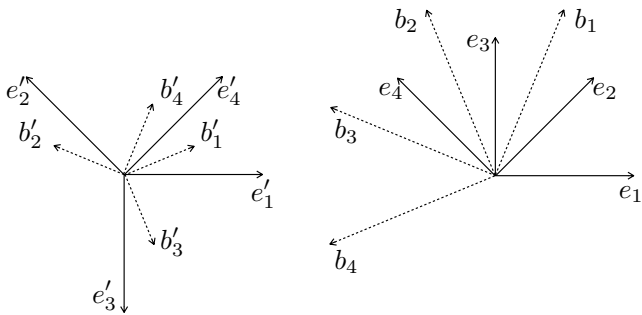}
\caption { Projections of the basis vectors $\varepsilon_n$ and of the BCC lattice vectors $\beta_n$  in the perpendicular plane $P'$ (left) and the real plane $P$ (right).}
\label{fig_vecs}
\end{figure}

The wave vectors of the laser beams, $\vec{k}_n$, can be regarded as projections in the $xy$ plane of orthogonal four-dimensional vectors $\vec{K}_n=K\vec{\varepsilon}_n$, where $\vec{\varepsilon}_n$ is an orthonormal basis set. The magnitude of each vector is $K=\sqrt{2} k$.
 It is well known that two dimensional lattices cannot possess eight-fold rotational symmetry. The four dimensional hypercubic lattice, $\mathbb{Z}^4$, however, does. The 4D space is, moreover, the direct sum of two orthogonal invariant planes  $P$ and $P'$, having an irrational orientation with respect to the standard basis. Introducing the two orthogonal projection operators $\boldsymbol{\pi}$ and $\boldsymbol{\pi}'=I-\boldsymbol{\pi}$, one can write
 $\vec{R}=(\vec{r},\vec{r}')$, where  $\vec{r}=\boldsymbol{\pi}(\vec{R})$ is the projection of a given point in $P$, and $\vec{r}'=\boldsymbol{\pi}'(\vec{R})$ is its projection in $P'$.
One can choose orthogonal bases $\{e_x,e_y\}$ in $P$ and $\{e'_x,e'_y\}$ in $P$ where the 
projections $e_n=\boldsymbol{\pi}(\varepsilon_n)$ and  $e'_n=\boldsymbol{\pi}'(\varepsilon_n)$ are shown in Fig. \ref{fig_vecs}. 

A given point $\vec{R}$, written $(R_1,R_2,R_3,R_4)$ in the standard 4d basis, can also be written $(\vec{r},\vec{r}')=(x,y,x',y')$ in the $\{e_x,e_y,e'_x,e'_y\}$ bases of 
$P$ (the ''physical space" or $xy$ plane, in which the atoms are located) and $P'$ (the ''perpendicular space" of the same system). These coordinates are related to the $\{R_n\}$ by a 4d rotation $\mathcal{R}$: 
\begin{align*}
\left[\begin{array}{l}
x\\
y\\
x'\\
y'
\end{array}\right] &=\mathcal{R}
\left[\begin{array}{l}
R_1\\
R_2\\
R_3\\
R_4
\end{array}\right]
=\frac 1 2
\left[\begin{array}{rrrr}
\sqrt{2} & 1 & 0 & -1\\
0 & 1 & \sqrt{2} & 1 \\
\sqrt{2} & -1 & 0 & 1 \\
0 & 1 & -\sqrt{2} & 1
\end{array}\right].
\left[\begin{array}{l}
R_1\\
R_2\\
R_3\\
R_4
\end{array}\right].
\end{align*}

If $\vec{K}=(\vec{k},\vec{k}')$ is another vector, the scalar product writes 
$\vec{K}.\vec{R}=\sum K_n R_n=\vec{k}.\vec{r}+\vec{k}'.\vec{r}'$ by orthogonality of $P$ and $P'$.  Fiinally, we introduce the BCC lattice $B$ obtained by adding the body centers to $\mathbb{Z}^4$. The four primitive lattice vectors of this BCC lattice, $\beta_n$, project onto  
the set $b_n$,and $b'_n$ in $P$ and $P'$ shown in Fig. \ref{fig_vecs}. One sees that they are turned by angles of $3\pi/8$ and $\pi/8$ with respect to the set $e_n$ and $e'_n$, exactly what one sees in the tilings of  Fig.\ref{one.fig}.

The optical potential Eq.\ref{Vxy.eq} can be obtained from a 4d periodic function, 
\begin{equation}
\mathcal{V}(\vec{R})=V_0\left[\sum_{n=1}^4 \cos(\vec{K}_n.\vec{R})\right]^2
\label{V4d.eq}
\end{equation}
 We let $\vec{K}_n=(\vec{k}_n,\vec{k}'_n)$ with projections $\vec{k}_n$ on $P$ and $\vec{k}'_n$ on $P'$. It is easy to see that $V(\vec{r})=\mathcal{V}(\vec{r},0)$. Notice that phases $\phi_n$ in Eq.\ref{Vxy.eq} are equivalent to a global 4d translation $\tau$
in Eq.\ref{V4d.eq}, such that $K \tau.\varepsilon_n=\phi_n$. Such translations yield OQ's belonging to the same "local isomorphism class" (same patterns of finite size with the same probability).  

\begin{figure}[!ht]
\centering
\includegraphics[width=140pt]{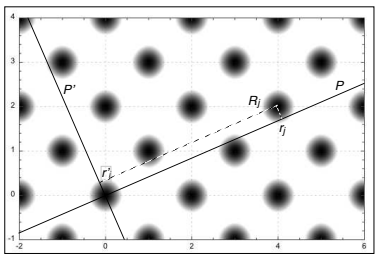}
\caption {Schematic of the domains where the condition $\mathcal{V}<V_c$ is satisfied. }
\label{fig_three}
\end{figure}

The maxima of $\mathcal{V}$ lie on the vertices of the BCC lattice $(\sqrt{2}\pi/k) B$, and the condition $\mathcal{V}(\vec{R})\leq V_c$ defines domains centered on lattice points as shown in Fig.\ref{fig_three}. If  the cutoff $V_c = (I_c/I_0)V_0$ is low enough, one can substitute the quadratic approximation  $\mathcal{V}(\vec{R})\approx V_0(16-8k^2(\vec{r}^2+\vec{r}'^2))$. The domains are then spheres of radius $\rho$ such that $8k^2\rho^2=\Delta V=V_c-16V_0$, while their projections on $P$ (or $P'$) are disks of area $D$ (or $D'$).  One can get the 4d coordinates $\vec{R}$ of points $\vec{r}$ where $V$ has a local minimum by using the inverse rotation $\mathcal{R}^t$. Such points are close to vertices of the body centered lattice, hence the connection with the octagonal tiling obtained by the cut-and-project method, as discussed below. We note that the optical quasicrystal thus obtained is robust under small changes in $V_c$.

\begin{figure}[!ht]
\centering
\includegraphics[width=120pt]{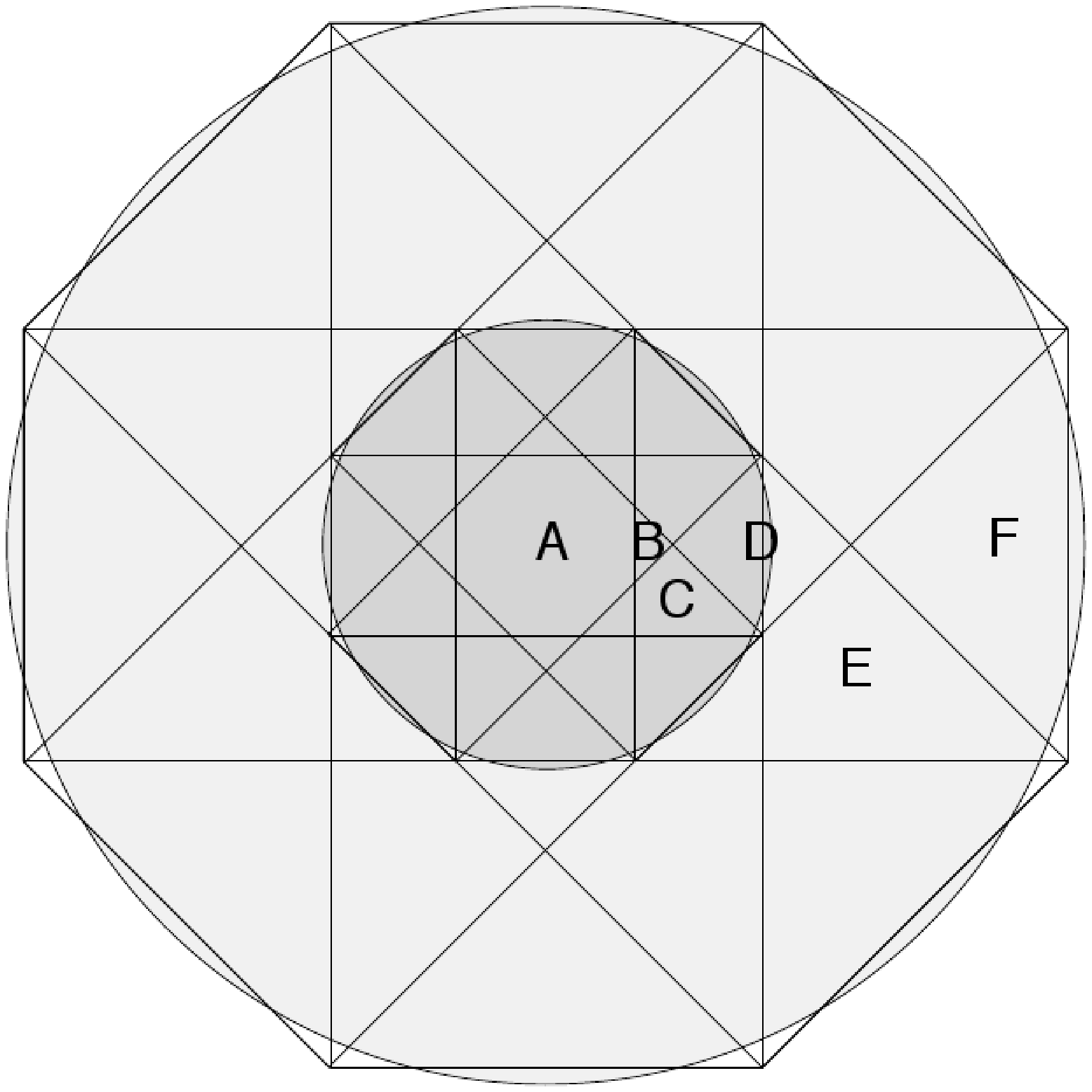}  \hskip 2cm
\includegraphics[width=120pt]{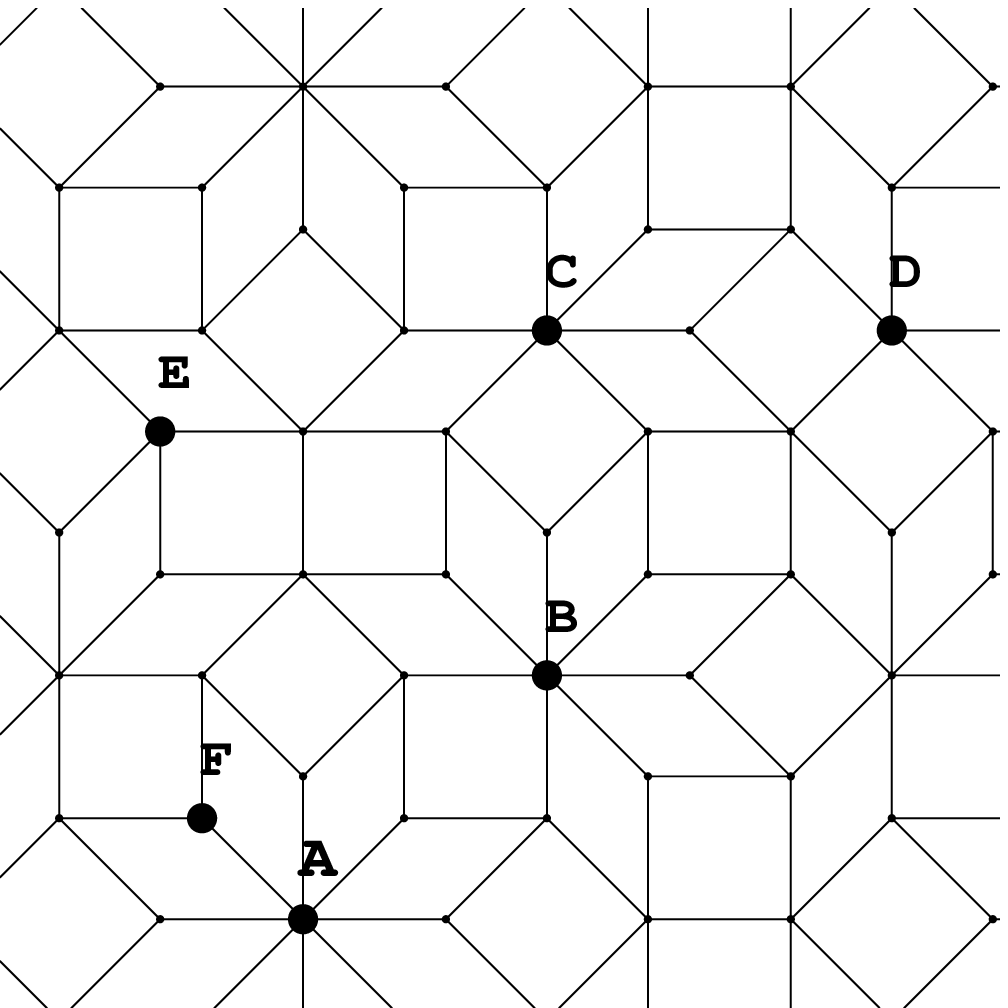} 
\caption{a) (top) The big octagon represents W, the OT selection window in $P'$. Subwindows for each  local environment (A,B,..) are shown., Circles represent the selection windows $D'$ for the OQ for the values $p=0$ and $p=1$.b) (bottom) A portion of the OT with the corresponding local environments.}
\label{fig_four}
\end{figure}

We turn next to the standard octagonal tiling. It is composed of all those points $r$, with $R=(r,r')$,  whose projections $r'$ in $P'$ fall within an octagonal window determined by the vectors $b'_n$ (see \cite{octagonal1,octagonal4} for the cut and project algorithm). The area of this selection window, shown in Fig.\ref{fig_four}, is $W=\frac{\sqrt{2}\pi^2}{k^2}$. We referred earlier to the inflation transformation of the octagonal tiling, whereby the edge length of the tiles in $P$  increase by a factor $\alpha$. It can be shown that, concomitantly, distances in $P'$ are reduced by the same factor.  Inflated octagonal tilings are therefore associated with selection windows which have an area of $W/\alpha^{2p}$ where $p=0,1,2,..$.  

As $V_c$ is varied, and provided it is low enough, a relationship with an octagonal tiling is expected if the windows $D'$ and $W$ (up to inflation) are similar in area. We require that the areas of the domains be the same, ensuring that the areal density of points is the same in the two structures, and this gives the condition 
\begin{equation}
\Delta V/V_0=8\sqrt{2}\pi\alpha^{-2p}
\label{cond.eq}
\end{equation}

For all values of $V_c$ which satisfy the condition Eq.\ref{cond.eq} one obtains a structure closely related to the standard octagonal tiling.  The selection windows $D'$ for different values of $p$ are shown in Fig.\ref{fig_four}, inside the octagonal selection window. The edges of the OQ have length $\ell=\frac{\sqrt{2}\pi}{k}\,|b_n|\alpha^p$. The smallest edge length obtainable by the method corresponds to the case $p=1$, when $\ell \approx 3.81 \lambda$. The OQ of Fig.\ref{one.fig} correspond to $p=2$ and $3$. Other values of the cutoff yield different structures, outside the scope of this paper.

\begin{figure}[!ht]
\centering
\includegraphics[width=120pt]{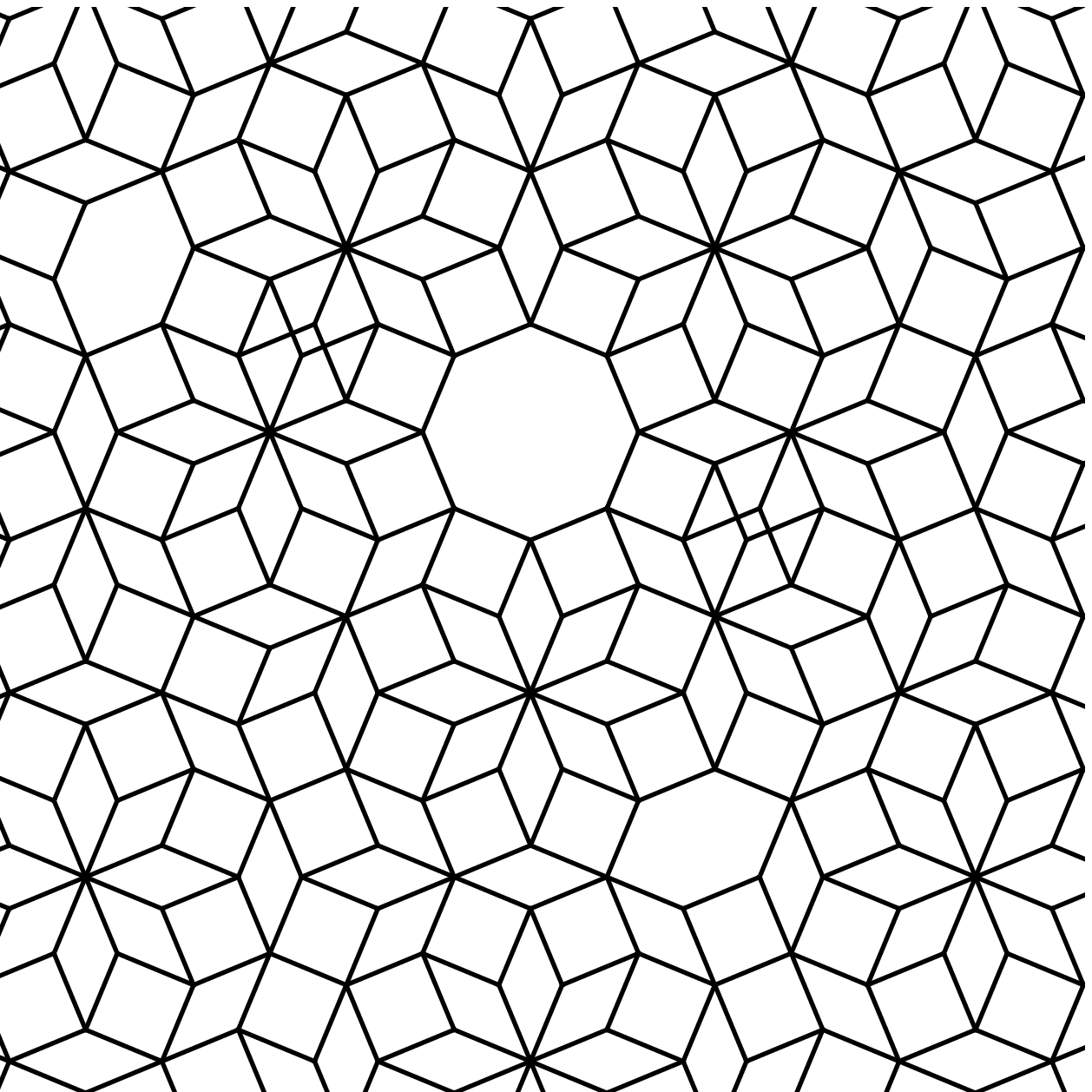}  \hskip 2cm
\includegraphics[width=140pt]{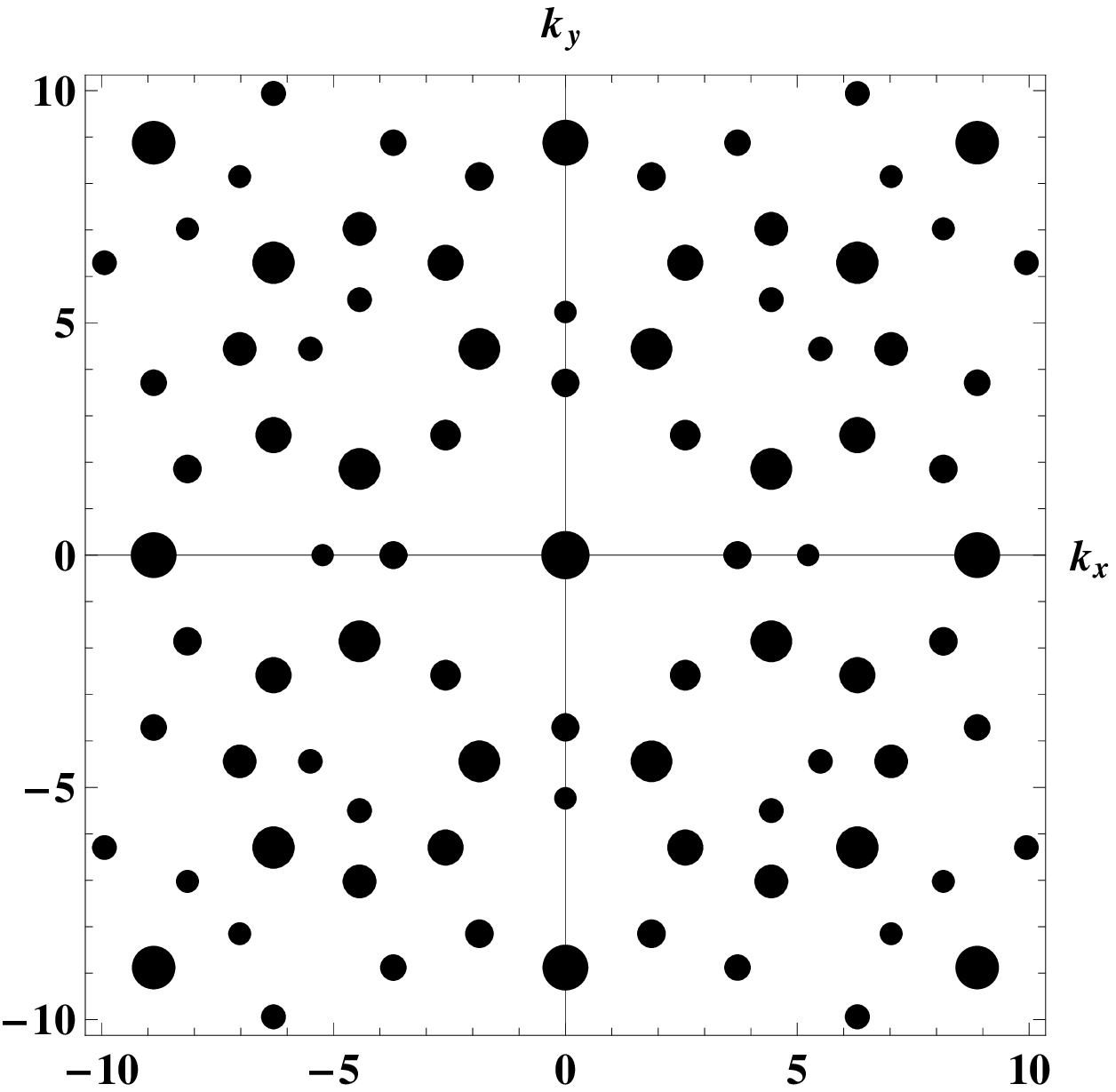}
\caption{a) (top)  A portion of an OQ : note empty hexagons and octagons, and close lying pairs of sites with overlapping bonds. b) (bottom) The structure factor for a sample OQ with 748 atoms (p=1). Intensities of the peaks are proportional to the area of the spots, $k_x$ and $k_y$ are given in units of $\lambda^{-1}$.}
\label{fig_OQ}
\end{figure}

\section{Structures and effective tight-binding models \label{sec_tbmodel}}
We now discuss the structure differences between the OT and the optical quasicrystal shown in Figs.\ref{fig_four}b) and \ref{fig_OQ}a). The differences between the two can be explained by the difference in shape of the domains of acceptance in $P'$: an almost circular disk for the former, and an octagon (of equal area) for the latter. The coordination number $z$ of any vertex of the OT is related to its position in $P'$, as shown in Fig. \ref{fig_four}a), where $z=8,7,...3$ correspond to points of the domains indicated by the letters A,B,...F.  When the windows of the OT and the OQ are overlapped, one sees that the differences arise in the peripheral region, namely the E and F domains. Thus, the OQ  i)  is missing some sites, whence the empty octagons and hexagons, and ii) it has some new sites which appear in close proximity to others, forming twin-pairs. Such pairs of sites are well-known in the literature on quasicrystals, as ``phason-flip"  conjugates \cite{hen}. Whereas in quasicrystals only one of the pair of sites is occupied, in the present case of the OQ,  $both$ sites are simultaneously occupied, leading to bonds that cross each other. These differences, which concern a small fraction of atomic positions, differentiate the OQ from the standard OT, which is a more homogeneous structure.  Adding higher harmonics in Eq.\ref{Vxy.eq} would ensure a better overlap of windows and thus a better coincidence of the octagonal tiling and the optical quasicrystal. Alternatively this could be achieved by introducing repulsive interactions, which would tend to favor homogeneity.

At low temperature, atoms occupy the lowest energy state of their wells. In this limit one can appropriately describe the optical quasicrystal using a Wannier basis set, localized on the sites. As a first approximation, one can use the set of harmonic oscillator ground state wavefunctions, approximating the exact potential locally by harmonic oscillators. In this basis the diagonal matrix elements of the Hamiltonian are $V_i $ which have small local variations due to the intensity field and the harmonic trap potential confining the atoms. The off-diagonal elements, $t_{ij}=-\langle i\vert H\vert j\rangle$ correspond to the amplitude of tunnelling between sites $i$ and $j$.

The simplest noninteracting model of particles in the OQ is thus described by a hopping Hamiltonian of the form
\begin{equation}
H = -\sum_{\langle i,j\rangle} t_{ij} (a^\dagger_{i} a_{j} + a^\dagger_{j} a_{i}) + \sum_{i=1}^N V_ia^\dagger_{i} a_{i},
\end{equation}
where the operator $a_{i} (a^\dag_i)$ annihilates(creates) a particle at site $i$ of the OQ, and sites are labeled $i,j=1,...N$ where $N$ is the total number of lattice sites. In the kinetic  term, it is sufficient to consider a small subset of hoppings between near neighbor sites $i$ and $j$. In the OQ, the smallest distances are $d_s$, the short diagonal of the rhombus, and $\ell$, the edge. There is in addition a shorter distance, $\delta=(\sqrt{2}-1)\ell$, between twin-pairs, but these are infrequent and can moreover be eliminated. The OQ differs from the OT in having regions of higher density (where twins occur) and of lower density (where empty hexagons and octagons occur). They result from imposing a rigid cutoff $I_c$. If one assumes a smooth cutoff, and one turns on a weak repulsive interaction between atoms, these defects would be energetically unfavorable.

\begin{figure}[!ht]
\centering
\includegraphics[width=150pt]{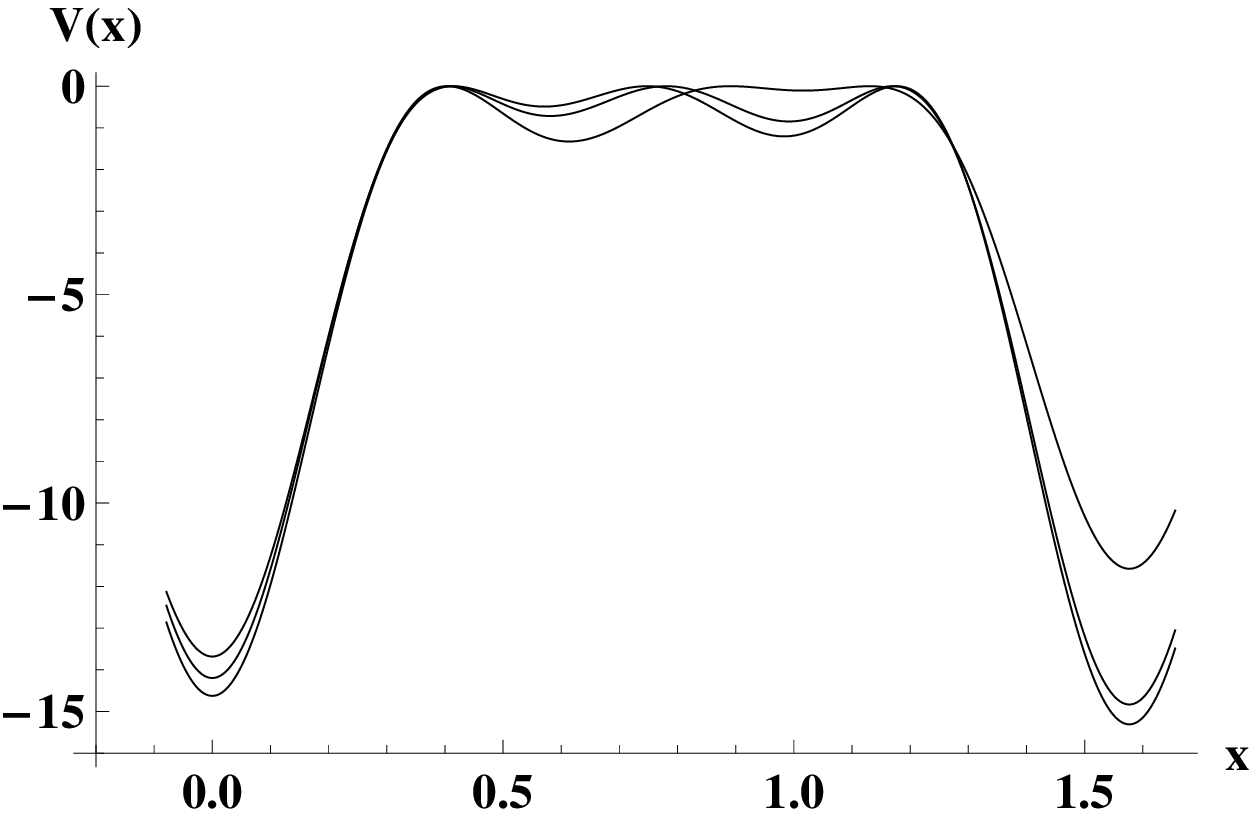} \hskip 1cm
\includegraphics[width=150pt]{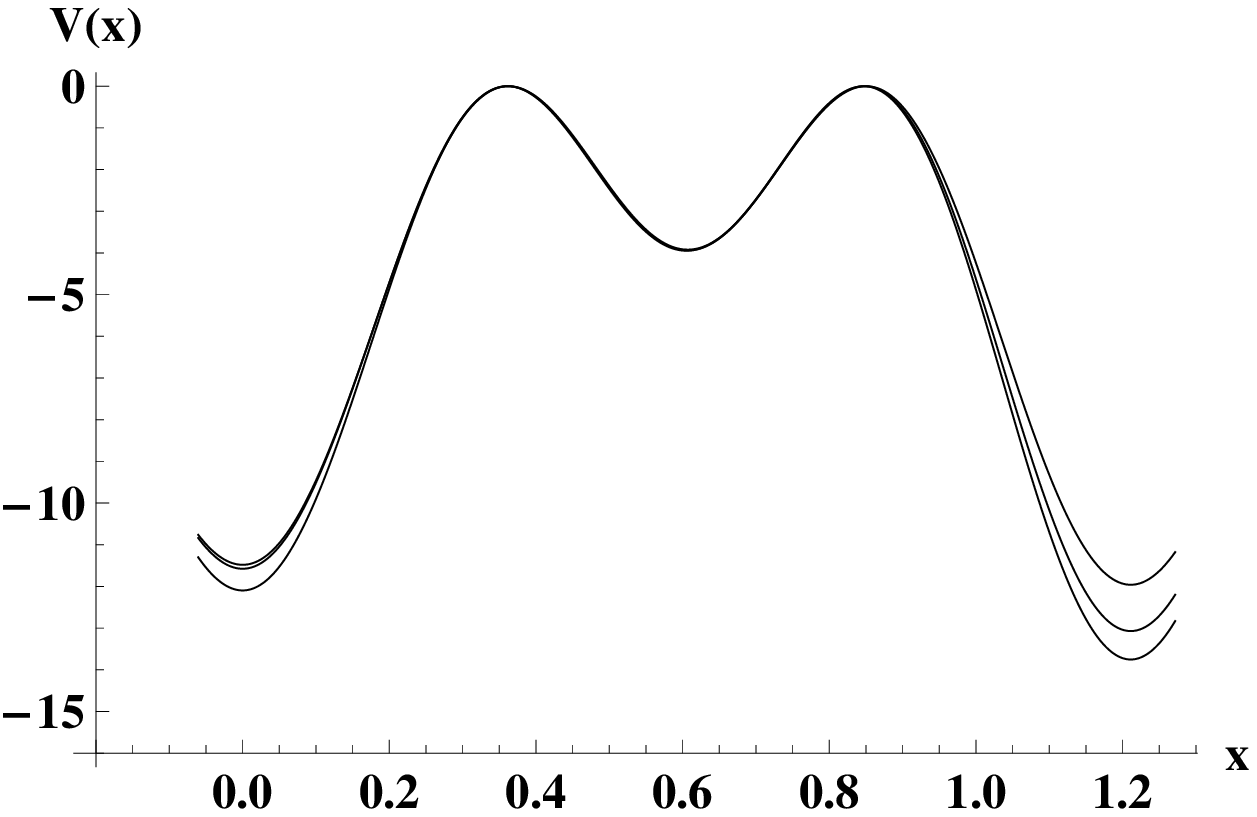} 
\caption{ Plots showing for several pairs the potential energy variation along edges  (top)  and across short diagonals (bottom) in the OQ for $p=1$. Distances in units of $\lambda$.}
\label{fig_potvar}
\end{figure}

Consider now the optical quasicrystal for $p=1$. Although details differ for each pair, there are strong similarities between the potential barriers seen by the atoms a) in the case of hopping  across small diagonals, and b) of hopping along edges, as illustrated in Fig.\ref{fig_potvar}.  Based on a WKB approach, we expect, to first approximation, that the typical hopping amplitude for edges, $t$, should be smaller than the typical hopping amplitude $t_s$ for diagonal hops  (as can be checked by numerically integrating the action along the two different pathways). It  is useful to recall results for the hopping amplitude for the periodic case, where atoms are trapped in potential wells of height $V$ due to a standing wave of wavelength $\lambda$. The hopping amplitude from one well to the other is  $\propto \exp(-cst \sqrt{V/E_r})$, where $E_r$ is the  recoil energy, $E_r = h^2/2m\lambda^2$, provided that $V\gg E_r$ \cite{bloch}. The prefactor, more difficult to determine, has been calculated, for example, in the case of a graphene-type structure \cite{miniatura,ibanez}, while for a more general case, a method is outlined in \cite{milnikov}. For the OQ for $p=2$, we find that the situation is reversed: the tunnelling amplitude is $larger$ for edge hopping, i.e. $t_s/t \ll 1$. The determination of the hopping amplitudes in the OQ for different $p$ is left for future work.

One expects that the tight-binding models on the OT and on the OQ will share certain features since both are invariant under inflation. Results have been obtained for the OT spectrum and wavefunctions \cite{sire1,ben}, for local densities of states \cite{localdos}, energy level statistics \cite{philmag}, for quantum dynamics \cite{ben2,moss2}, effect of Hubbard interaction \cite{jagsch}, and the Heisenberg spin limit \cite{wess,jag}. The optical quasicrystal is expected to share many of the properties found.  Although the form of the density of states will be nonuniversal, other characteristics such as singular local response functions, multifractal spectrum and wavefunctions, anomalous diffusion, and self similar magnetic states should be found. In particular, in the cold atom quasicrystal, it should be possible to create a wavepacket in a specified initial state, study its subsequent mean square displacement $\langle r^2\rangle \sim t^{2\beta}$ with time and extract $\beta$. For the OT, when $V_i=0$, the exponent $\beta$ is larger than $\frac{1}{2}$ , which is the value for normal diffusion, and it is nonuniversal, depending, for example, on the initial position \cite{ben2}.

\section{Summary and discussion}
We have discussed the 8-fold optical quasicrystal obtained by trapping atoms in the quasiperiodic potential landscape created by four standing laser waves.  We have discussed the structures obtained for a particular series of values of the cutoff, and related them to the well-known octagonal tiling using a four dimensional description.  By changing the laser intensity at a fixed temperature, one can pass from a given OQ to an ``inflated" OQ of bigger edge length. The tight-binding approximation for such an optical quasicrystal was discussed qualitatively. One can expect that some of the experimental difficulties of realizing such an optical quasicrystal will concern laser alignment, phase stabilization and effective trapping of atoms. This optical quasicrystal would, if realized, provide an ideal system in which to study the quantum physics of quasiperiodic structures. It would be interesting, in particular, to experimentally observe the superdiffusion of wavepackets in the quasicrystal. 

\acknowledgments
We would like to thank Christoph Weitenberg (LKB, Paris), Bess Fang (Institute of Optics, Palaiseau), Monika Aidelsburger (LMU, M\"unchen), J.-M. Luck and J-.F. Sadoc for useful discussions.


\begin{thebibliography}{26}

\bibitem{bloch}
Immanuel Bloch, Jean Dalibard, and Wilhelm Zwerger.
\newblock Many-body physics with ultracold gases.
\newblock {\em Rev. Mod. Phys.}, 80:885, 2008.

\bibitem{beenker}
F.P.M. Beenker.
\newblock {\em Algebric theory of non periodic tilings of the plane by two
  simple building blocks: a square and a rhombus,TH Report 82-WSK-04}.
\newblock Technische Hogeschool, Eindhoven, 1982.

\bibitem{grun}
B.~Gruenbaum and G.~C. Shephard.
\newblock {\em Tilings and Patterns}.
\newblock San Francisco, 1987.

\bibitem{grimm}
R.~Grimm, M.~Weidemueller, and Y.B. Ovchinnikov.
\newblock Optical dipole traps for neutral atoms.
\newblock {\em Adv. At. Mol. Opt. Phys.}, 42:95, 2000.

\bibitem{bohr}
H.~Bohr.
\newblock {\em Almost-periodic functions}.
\newblock New York, 1947.

\bibitem{besic}
A.S. Besicovitch.
\newblock {\em Almost periodic functions}.
\newblock Dover, Cambridge, 1954.

\bibitem{deiss}
B.~Deissler, E.~Lucioni, M.~Modugno, G.~Roati, L.~Tanzi, M.~Zaccanti,
  M.~Inguscio, and G.~Modugno.
\newblock Correlation functions of weakly interacting bosons in a disordered
  lattice.
\newblock {\em New Journal of Physics}, 13:023020, 2011.

\bibitem{guidoni}
L.~Guidoni, C.~Trich\'e, P.~Verkerk, and G.~Grynberg.
\newblock Quasiperiodic optical lattices.
\newblock {\em Physical Review Letters}, 79:3363, 1997.

\bibitem{sanchez}
L.~Sanchez-Palencia and L.~Santos.
\newblock Bose-einstein condensates in optical quasicrystal lattices.
\newblock {\em Physical Review A}, 72:053607, 2005.

\bibitem{cetoli}
Alberto Cetoli and Emil Lundh.
\newblock Towards a bose glass transition in an optical penrose lattice.
\newblock {\em ArXiv 1107.3062}.

\bibitem{bechinger}
J.~Mikhael, G.~Gera, T.~Bohlein, and C.~Bechinger.
\newblock Phase behavior of colloidal monolayers in quasiperiodic light fields.
\newblock {\em Soft Matter}, 7:1352, 2011.

\bibitem{octagonal1}
P.~J. Steinhardt and S.~Ostlund.
\newblock {\em The Physics of Quasicrystals}.
\newblock Singapore, 1987.

\bibitem{octagonal4}
J.-B.Suck, M.~Schreiber, and P.~Haussler.
\newblock {\em Quasicrystals}, volume~55.
\newblock Berlin, 2002.

\bibitem{hen}
P.J. Steinhardt and D.P. Vincenzo, editors.
\newblock {\em Quasicrystals:The State of the Art}.
\newblock World Scientific, 1991.

\bibitem{miniatura}
Kean~Loon Lee, Benoit Gr\'emaud, Rui Han, Berthold-Georg Englert, and Christian
  Miniatura.
\newblock Ultracold fermions in a graphene-type optical lattice.
\newblock {\em Physical Review A}, 80:043411, 2009.

\bibitem{ibanez}
Julen Ibanez-Azpiroz, Asier Eiguren, Aitor Bergara, Giulio Pettini, and Michele
  Modugno.
\newblock Tight-binding models for ultracold atoms in honeycomb optical
  lattices.
\newblock {\em Phys. Rev. A}, 87:011602, 2013.

\bibitem{milnikov}
V.~Mil'nikov and Hiroki Nakamura.
\newblock {\em J. Chem. Phys.}, 115:6881, 2001.

\bibitem{sire1}
C.~Sire and J.~Bellissard.
\newblock {\em Europhys. Lett.}, 11:439, 1990.

\bibitem{ben}
Vincenzo~G. Benza and Cl\'ement Sire.
\newblock {\em Phys. Rev. B}, 44:10343, 1991.

\bibitem{localdos}
A.~Jagannathan.
\newblock {\em J. de Physique}, 55, 1994.

\bibitem{philmag}
A.~Jagannathan and F.~Pi\'echon.
\newblock {\em Philosophical Magazine}, 87:2389, 2006.

\bibitem{ben2}
B.~Passaro, C.Sire, and V.G. Benza.
\newblock {\em Phys. Rev. B}, 46:13751, 1992.

\bibitem{moss2}
J.X.Zhong and R.~Mosseri.
\newblock {\em J. Phys. I (France)}, 4:1513, 1994.

\bibitem{jagsch}
A.~Jagannathan and H.~J. Schulz.
\newblock {\em Phys. Rev. B}, 55:8045, 1997.

\bibitem{wess}
S.~Wessel, A.~Jagannathan, and S.~Haas.
\newblock {\em Phys. Rev. Lett.}, 90:177205, 2002.

\bibitem{jag}
A.~Jagannathan.
\newblock {\em Phys. Rev. Lett.}, 92:047202, 2004.

\end{thebibliography}
\end{document}